\documentclass[fleqn,twoside]{article}
\usepackage[centertags]{amsmath}
\usepackage{espcrc2}
\usepackage{graphicx}
\usepackage[figuresright]{rotating}
\usepackage{tabularx}

\newcommand{\be}{\begin{equation}}
\newcommand{\ee}{\end{equation}}

\title{External Fields and Color Confinement}

\author{P. Cea\address{Dept. of Physics  and INFN - Bari - Italy},
        L. Cosmai\address{INFN - Bari - Italy}}

\begin{document}

\begin{abstract}
U(1), SU(2), and SU(3) lattice gauge theories in presence of
external fields are investigated both in (3+1) and (2+1) dimensions.
The free energy of gauge systems has been measured.
While the phase transition in compact U(1) is not influenced by the strength
of an external constant magnetic field, the deconfinement temperature
for SU(2) and SU(3) gauge systems in a constant abelian chromomagnetic
field decreases when the strength of the applied field increases.
The dependence of the deconfinement temperature on the strength of an external
constant chromomagnetic field seems to be a peculiar feature of
non abelian gauge theories.
\vspace{1pc}
\end{abstract}

\maketitle


\section{Introduction}

Color confinement is still a puzzling problem not withstanding the large mess
of numerical investigations aimed to understand the nature of the QCD vacuum.
Therefore we feel that it is a good task to explore new paths that possibly may suggest new ideas
towards understanding confinement.

In a seminal paper~\cite{Feynman:1981ss}   R. P. Feynman argued that
in non abelian gauge theories long range correlation between gluonic degrees of
freedom can destroy confinement.
This suggested us that a uniform constant background field (that in this case
should restore long range correlations between gluonic degrees of freedom) could influence
the deconfinement temperature for non abelian gauge theories.

In order to investigate vacuum structure of lattice gauge theories at zero temperature
a lattice gauge invariant effective action $\Gamma[\vec{A}^{\text{ext}}]$ for an external
background field $\vec{A}^{\text{ext}}$
was introduced in Refs.~\cite{Cea:1997ff,Cea:1999gn}.
For a gauge theory at finite temperature $T=1/(a L_t)$
in presence of an external background field, the relevant quantity is
the free energy functional defined as (for more details and discussions,
see Refs.~\cite{Cea:2002wx,Cea:2004ux})
\be
\label{freeenergy}
{\mathcal{F}}[\vec{A}^{\text{ext}}] = -\frac{1}{L_t} \ln
\left\{
\frac{{\mathcal{Z_T}}[\vec{A}^{\text{ext}}]}{{\mathcal{Z_T}}[0]}
\right\} \; .
\ee
${\mathcal{Z_T}}[\vec{A}^{\text{ext}}]$ is the thermal partition
functional~\cite{Gross:1981br}
in presence of the background field $\vec{A}^{\text{ext}}$, and is defined as
\be
\label{ZetaTnew}
\mathcal{Z}_T \left[ \vec{A}^{\text{ext}} \right]
= \int_{U_k(\vec{x},L_t)=U_k(\vec{x},0)=U^{\text{ext}}_k(\vec{x})}
\mathcal{D}U \, e^{-S_W}   \,.
\ee
In Eq.~(\ref{ZetaTnew}) the spatial links belonging
to the time slice $x_t=0$ are constrained to the value of the external background field,
the temporal links are not constrained.

%
%
\section{Deconfinement Temperature vs. $gH$}

We study vacuum dynamics for U(1), SU(2), and SU(3) lattice
gauge theories under the influence of an
Abelian chromomagnetic background field.
In our previous studies we found that SU(3) deconfinement temperature
depends on the strength of an applied external constant
Abelian chromomagnetic field~\cite{Cea:2002wx}.
We would like to corroborate our findings with further investigations,
in particular we would like to ascertain if the dependence of the deconfinement
temperature on the strength of an applied external
constant Abelian chromomagnetic field is a peculiar feature of non abelian gauge theories.

In the continuum a static constant Abelian chromomagnetic field is given by:
\be
\label{field}
\vec{A}^{\mathrm{ext}}_a(\vec{x}) =
\vec{A}^{\mathrm{ext}}(\vec{x}) \delta_{a,3} \,, \quad
A^{\mathrm{ext}}_k(\vec{x}) =  \delta_{k,2} x_1 H \,.
\ee
On a lattice with hypertoroidal geometry the magnetic field turns out to be quantized
(i.e.: $a^2 g H/2 = 2 \pi/L_x$, $n_{\mathrm{ext}}$  integer).
The free energy $F[\vec{A}^{\mathrm{ext}}]$
is proportional to spatial
volume $V=L_s^3$ and the relevant quantity is the density
$f[\vec{A}^{\mathrm{ext}}]$ of
free energy.
We evaluate the $\beta$-derivative of $f[\vec{A}^{\mathrm{ext}}]$ at fixed
external field strength $gH$
($\Omega = L_s^3 \times L_t$):
\be
\label{deriv}
\begin{split}
f^{\prime}[\vec{A}^{\mathrm{ext}}]   & = \left \langle
\frac{1}{\Omega} \sum_{x,\mu < \nu}
\frac{1}{3} \,  \text{Re}\, {\text{Tr}}\, U_{\mu\nu}(x) \right\rangle_0  \\
& - \left\langle \frac{1}{\Omega} \sum_{x,\mu< \nu} \frac{1}{3} \,
\text{Re} \, {\text{Tr}} \, U_{\mu\nu}(x)
\right\rangle_{\vec{A}^{\mathrm{ext}}} \,,
\end{split}
\ee
where the subscripts on the averages indicate the value of the external field.

As is well known pure SU(3) gauge system undergoes a deconfinement
phase transition by increasing temperature.
We can evaluate the critical temperature $T_c$ by locating the peak of
$f^{\prime}[\vec{A}^{\mathrm{ext}}]$ as a
function of $\beta$ for different lattice temporal sizes $L_t$.
We vary the strength of the applied external
Abelian chromomagnetic background field to study quantitatively
the dependence of $T_c$ on $gH$.
\begin{figure}[t]
\begin{center}
\includegraphics[width=0.4\textwidth,clip]{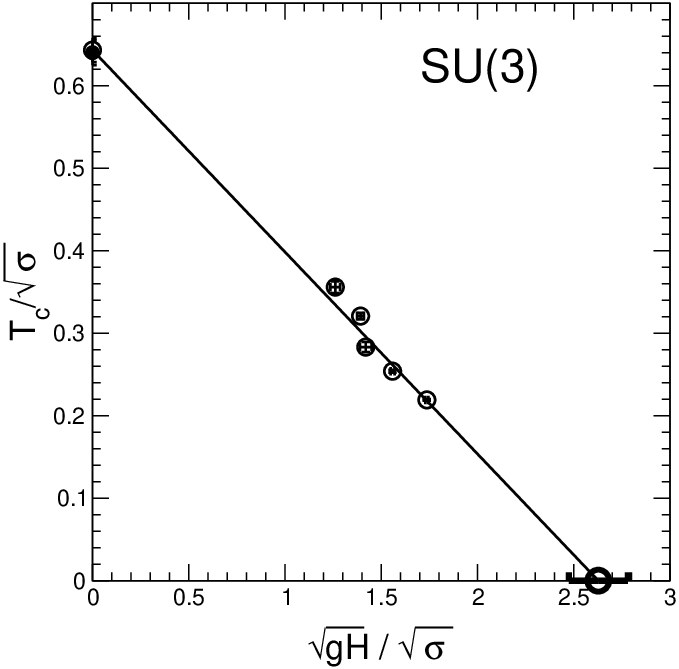}
\vspace{-1.1cm}
\caption{SU(3) in (3+1) dimensions.
$T_c/\sqrt{\sigma}$ vs. the applied field strength in units of string tension.
}
\end{center}
\end{figure}
\begin{figure}[t]
\begin{center}
\includegraphics[width=0.4\textwidth,clip]{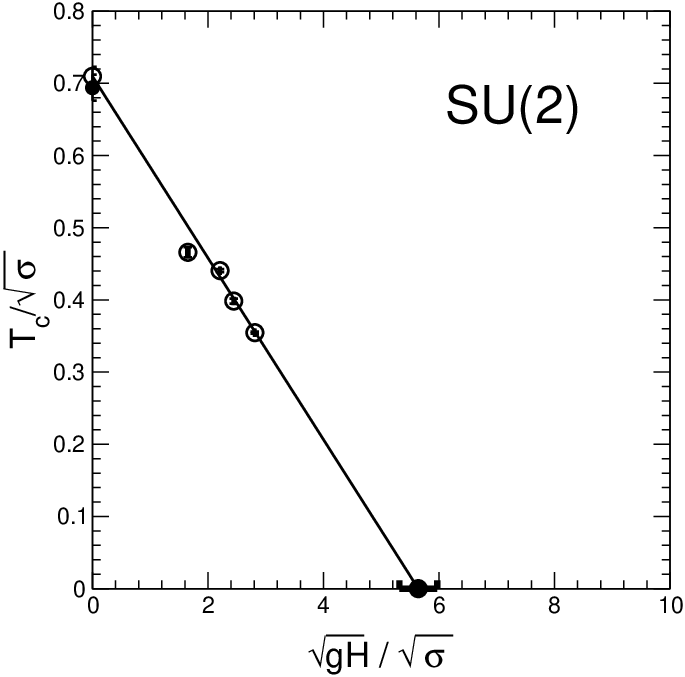}
\vspace{-1.1cm}
\caption{SU(2) in (3+1) dimensions.
$T_c/\sqrt{\sigma}$ vs. the applied field strength in units of string tension.
}
\end{center}
\end{figure}
To extract the critical field strength in physical units
we compute $T_c/\sqrt{\sigma}$ vs. $\sqrt{gH}/\sqrt{\sigma}$,
where the string tension at $\beta^*(L_t=8,n_{\text{ext}})$
is obtained from Eq.~(4.4) in Ref.~\cite{Edwards:1998xf}.
We obtain that the critical temperature decreases by increasing
the external Abelian chromomagnetic field.
If the magnetic length
$a_H \sim 1/\sqrt{gH}$ is the only relevant
scale of the problem for dimensional reasons one expect that
$T_c/\sqrt{\sigma} \sim \sqrt{gH}/\sqrt{\sigma}$.
Indeed we get a good linear fit to our data with the
critical field
$\sqrt{gH_c}/{\sqrt{\sigma}} = 2.63  \pm 0.15$ (see Fig.~1).

An analogous result has been found for SU(2) l.g.t., where
the critical coupling $\beta^*(L_t,n_{\text{ext}})$ has
been evaluated on a  $64^3 \times 8$ lattice,
and the string tension is obtained
according to Ref.~\cite{Teper:1998kw}.
In this case the   critical field is
$\sqrt{gH_c}/{\sqrt{\sigma}} = 5.33  \pm 0.33$ (see Fig.~2).
It is worthwhile to note that, using an effective approach within dual superconductor picture,
it was suggested~\cite{Chernodub:2002we}
that $gH_c/\sigma=1$ for SU(2) and $gH_c/\sigma=3/4$ for SU(3).

To ascertain if the effect we have found is peculiar of non Abelian
gauge theories we have simulated U(1) lattice gauge theory at zero temperature
in order to test a possible dependence of the confinement-Coulomb phase
transition on the strength of an applied constant magnetic field.
Indeed the critical coupling does not depend on the applied magnetic field strength,
in agreement with Ref.~\cite{Chernodub:2001mg} (see Fig.~3).
\begin{figure}[t]
\begin{center}
\includegraphics[width=0.4\textwidth,clip]{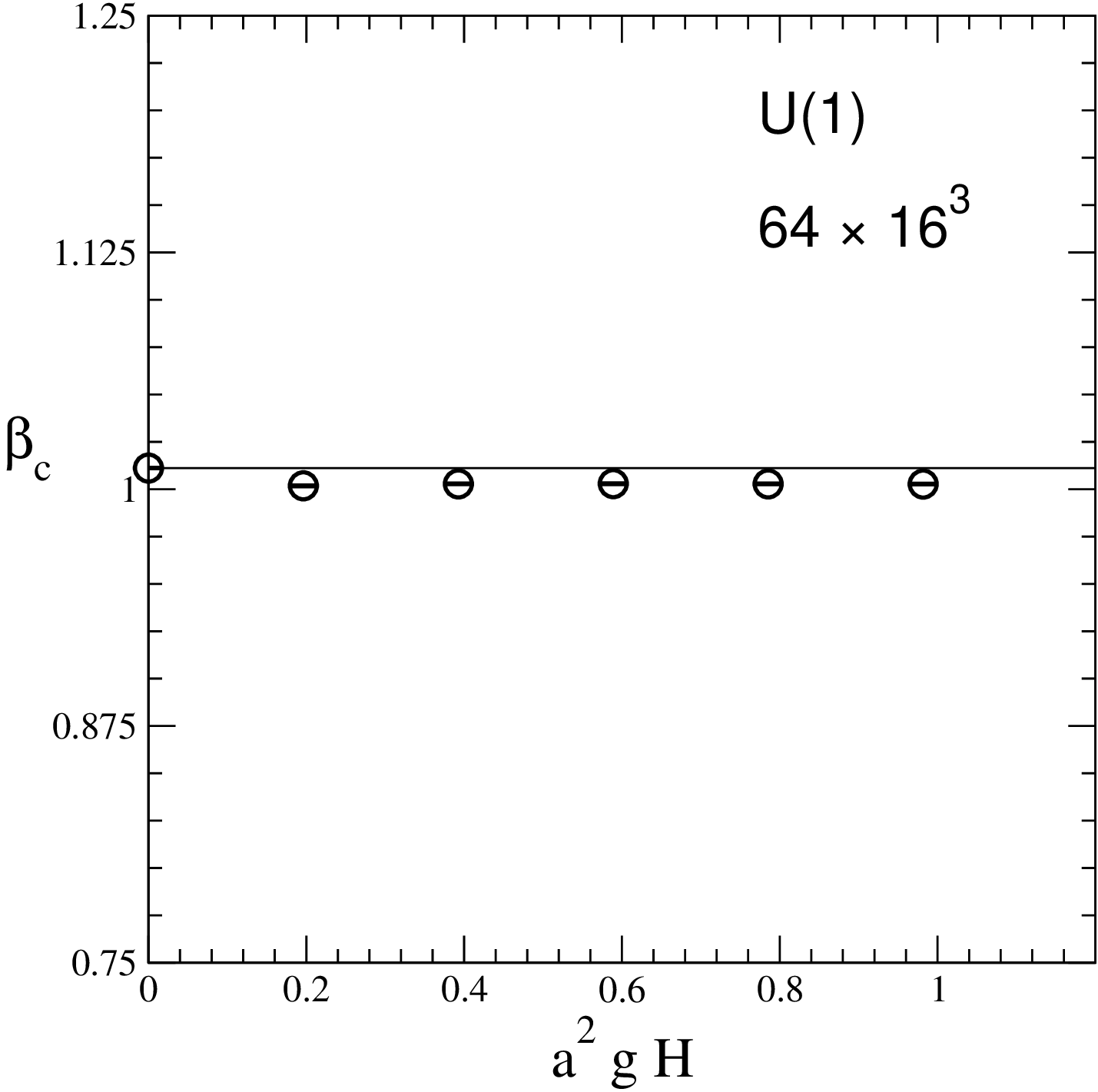}
\vspace{-1.1cm}
\caption{U(1) in 4 dimensions.
The critical coupling $\beta_c$ vs. the applied field strength in lattice units.
}
\end{center}
\end{figure}

Finally, to see if this {\em color Meissner effect} is a generic feature
of non Abelian gauge theories, we also consider gauge systems in (2+1) dimensions.
Gauge theories in (2+1) dimensions possess a dimensionful coupling, namely
$g^2$ has dimension of mass and so provides a physical scale.


In (2+1) dimensions the chromomagnetic field $H^a$
is a (pseudo)scalar
\be
\label{B3dim}
H^a = \frac{1}{2} \varepsilon_{ij} F^a_{ij} = F^a_{12}   \,.
\ee
For SU(3) gauge theory a constant abelian chromomagnetic field $H^3$
can be obtained with
\be
\label{3dlinks}
\begin{split}
& U^{\mathrm{ext}}_1(\vec{x}) = {\mathbf{1}} \,,
\\
& U^{\mathrm{ext}}_2(\vec{x}) =
\begin{bmatrix}
\exp(i \frac {g H x_1} {2})  & 0 & 0 \\ 0 &  \exp(- i \frac {g H
x_1} {2}) & 0
\\ 0 & 0 & 1
\end{bmatrix}
\end{split}
\ee
Again we find that $\frac{T_c}{\sqrt{\sigma}}$ depends linearly
on the applied field strength. A preliminary estimate, with $L_t=4$, gives
$\frac{T_c(0)}{\sqrt{\sigma}}= 1.073 (87)$, $\frac{gH_c}{\sqrt{\sigma}} = 5.5 \pm 3.7$, where the string tension has been taken from Ref.~\cite{Teper:1998te}.
The critical temperature value from Ref.~\cite{Engels:1997dz}
is $\frac{T_c(0)}{\sqrt{\sigma}}=0.972(10)$.
%
%


\section{Conclusions}
We have investigated U(1) and SU(2) in (3+1) dimensions,
and SU(3)  both in (3+1) and (2+1) dimensions.
Our numerical simulations were performed on APE machines in Bari.
For non abelian gauge theories we found that there is a critical field
$gH_c$ such that for $gH > gH_c$ the gauge systems are in the deconfined phase.
Such an effect is generic for non Abelian gauge theories and it is not shared by U(1) gauge theory.

As argued in Ref.~\cite{Feynman:1981ss} gauge invariance of the confining vacuum disorders
the system in such a way that there are not long range color correlations.
On the other hand strong enough chromomagnetic fields do introduce long range color correlations
such that the system gets deconfined. In conclusion it is worthwhile to observe that in general
the existence of a critical chromomagnetic field could be explained if
the confining vacuum behaves as a condensate of color charged fields whose mass is proportional
to the inverse of the magnetic length.

\end{document}